\documentclass{svproc}
\usepackage{graphicx}
\usepackage{multirow}
\usepackage{amsmath,amssymb,amsfonts}
\usepackage{mathrsfs}
\usepackage[title]{appendix}
\usepackage{xcolor}
\usepackage{textcomp}
\usepackage{manyfoot}
\usepackage{booktabs}
\usepackage{algorithm}
\usepackage{algorithmicx}
\usepackage{algpseudocode}
\usepackage{listings}
\usepackage{url}
\usepackage{hyperref}
\usepackage{sidecap}

\begin{document}
\mainmatter
\title{Updating the Complex Systems Keyword Diagram Using Collective Feedback and Latest Literature Data}
\titlerunning{Updating the Complex Systems Keyword Diagram}
\author{Hiroki Sayama\inst{1,2,3}}
\authorrunning{Hiroki Sayama}
\institute{Binghamton Center of Complex Systems, Binghamton University\\
State University of New York, Binghamton, NY, USA
\and
Waseda Innovation Lab, Waseda University, Tokyo, Japan
\and
Max Planck Institute for the Physics of Complex Systems, Dresden, Germany\\
\email{sayama@binghamton.edu} \quad
\texttt{https://bingdev.binghamton.edu/sayama/}
}

\maketitle

\begin{abstract}
The complex systems keyword diagram generated by the author in 2010 has been used widely in a variety of educational and outreach purposes, but it definitely needs a major update and reorganization. This short paper reports our recent attempt to update the keyword diagram using information collected from the following multiple sources: (a) collective feedback posted on social media, (b) recent reference books on complex systems and network science, (c) online resources on complex systems, and (d) keyword search hits obtained using OpenAlex, an open-access bibliographic catalogue of scientific publications. The data (a), (b) and (c) were used to incorporate the research community's and other public communities' perceptions of the relevant topics, whereas the data (d) was used to obtain more objective measurements of the keywords' relevance and associations from publications made in complex systems science. Results revealed differences and overlaps between public perception and actual usage of keywords in publications on complex systems. Four topical communities were obtained from the keyword association network, although they were highly intertwined with each other. We hope that the resulting network visualization of complex systems keywords provides a more up-to-date, accurate topic map of the field of complex systems as of today.
\keywords{complex systems, keyword associations, network visualization, topic communities, OpenAlex}
\end{abstract}

\section{Introduction}

\begin{SCfigure}[][t]
\centering
\includegraphics[width=0.6\columnwidth]{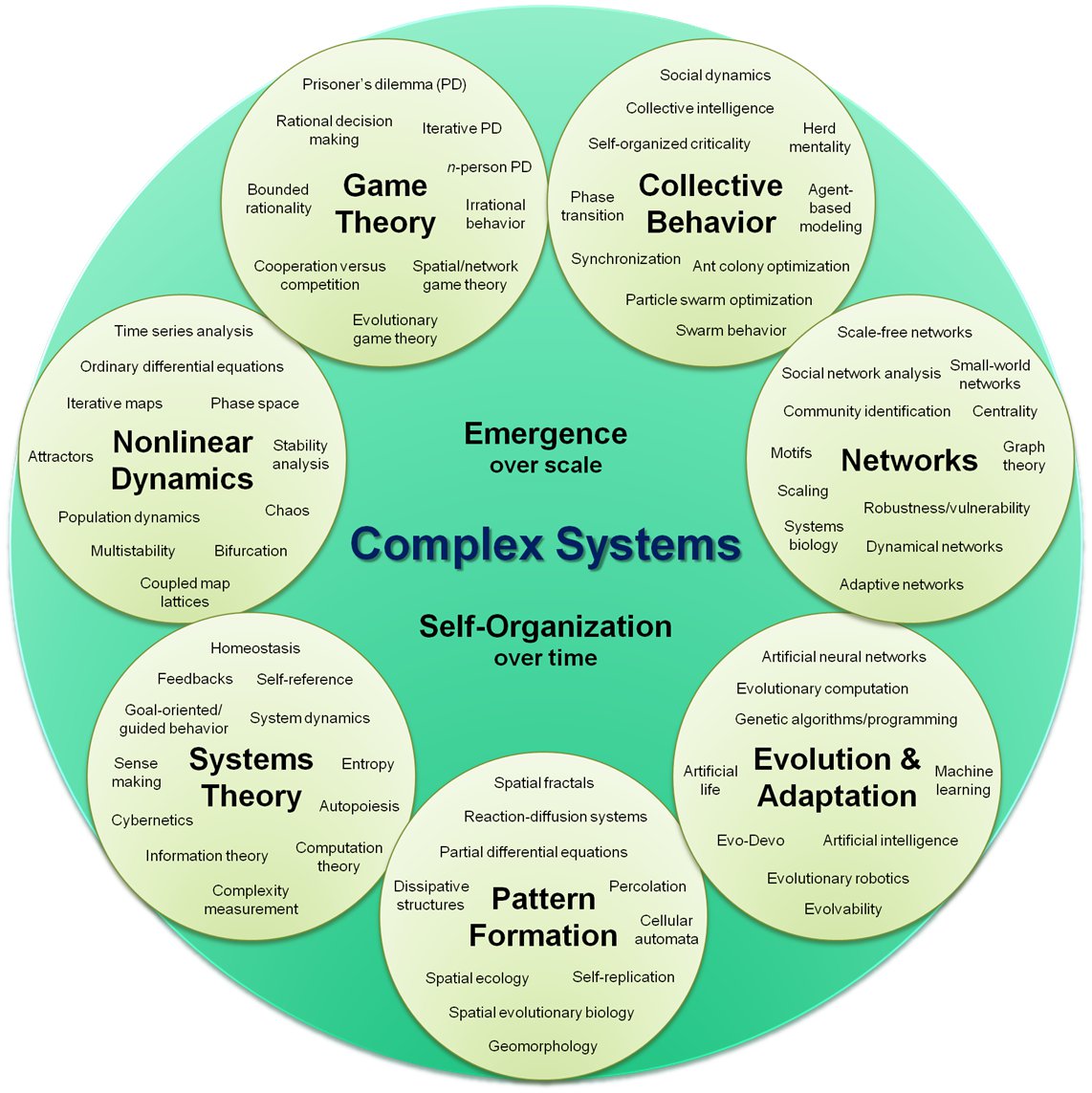}
\caption{The original complex systems keyword diagram created by the author in 2010 \cite{ref1}. This diagram was created manually according to the author's own personal views and subjective selection and categorization of topics back then.}
\label{fig:original}
\end{SCfigure}

The field of complex systems science covers a broad range of interdisciplinary topics that are often difficult to organize, making it challenging for newcomers to have a grasp of the whole picture of the field. To address this problem, the author created a conceptual organizational map of complex systems keywords in 2010 (Fig.~\ref{fig:original}) \cite{ref1}, which organized the topics into seven different groups  (nonlinear dynamics, systems theory, game theory, pattern formation, evolution \& adaptation, collective behavior, and networks). This keyword diagram has been used widely in a variety of educational and outreach purposes and has contributed to the dissemination and popularization of complex systems related concepts to a broader audience. 

More than a decade has passed since the creation of the original diagram, during which complex systems science experienced many new developments, including new network modeling frameworks (e.g., temporal networks, multilayer networks) and rapid adoption of AI, NLP, and other data-driven machine learning approaches. Moreover, the original diagram had a fundamental limitation as it was created based on the author's own personal views and subjective selection and categorization of topics. These issues call for a major update and reorganization of the complex systems keyword diagram, using a more objective, data-driven methodology where keywords and their relatedness are represented in a more quantitative weighted network. This short paper reports our recent attempt to update the keyword diagram using collective feedback from the community and the latest literature data.

\section{Data and Methods}

In this project, multiple data sources were used to collect a wide variety of keywords relevant to complex systems in both academic and non-academic uses. More specifically, the following four data sources were used:
\begin{description}
\item[(a) Collective feedback posted on several social media.] The author posted an inquiry to 
Twitter/X, Facebook, LinkedIn, and Mastodon on December 21, 2022, asking the community what kind of keywords, concepts, topics, and/or research areas should be added to the complex systems keyword diagram \cite{ref2}. This resulted in significant engagements on those online platforms (e.g., the post on Twitter/X gained 235k views, 1.6k likes, 605 bookmarks and 350 retweets) and more than 130 responses were received as a result.
\item[(b) Several recent reference books.] Topics were manually sampled from the tables of contents of several recent reference books on complex systems and network science \cite{ref3,ref4,ref5,ref6,ref7,ref8,ref9}, most of which were published within the last decade.
\item[(c) Several online resources on complex systems.] Similarly, topics were also manually sampled from several well-known online educational resources on complex systems \cite{ref10,ref11,ref12,ref13}.
\item[(d) Keyword search hits obtained using OpenAlex.] Keyword search hits in scientific literature were obtained using the API of OpenAlex \cite{ref14}, an open-access bibliographic catalogue of scientific publications, for individual keywords identified above, as well as {\em pairs} of keywords to quantitatively assess their relatedness.
\end{description}

The data (a), (b) and (c) were used to incorporate the research community's and other public communities' perceptions of the relevant topics, whereas the data (d) was used to obtain more objective measurements of the keywords' relevance and associations from publications made in complex systems science. In (a)-(c), the relevant keywords were manually selected and curated by the author. In (d), the overall relevance of each keyword to complex systems was characterized by its total search hit multiplied by its {\em visibility boost} \cite{ref15} when it was searched together with ``complex systems'' OR ``complexity science''. Namely, if the relevance score of a keyword is high, that implies that the keyword was more manifested in the context of complex systems or complexity science (see \cite{ref15} for details).

One problem faced in collecting the data (d) above was that many combinations of keywords resulted in superficial large search hits that might not correctly capture their actual scientific relatedness, especially when both of them were common words/phrases. To avoid such misleading measurements and improve the quality of network reconstruction, the entire relevant keyword list and possible associations of the pairs of keywords were manually reviewed and curated by the author from the viewpoint of complex systems science and engineering. The keyword association network was then constructed by assigning edge weights between a pair of associated keywords according to the OpenAlex search hit of the pair (searched together with ``complex systems'' OR ``complexity science'') multiplied by both keywords' relevance scores.

Wolfram Research Mathematica was used for all of the data collection, analysis and visualization for this project. All the collected data, the keyword association list, and the program codes that were used for data collection, analysis and visualization are publicly available from the project's GitHub page \cite{project-github}.

\section{Results}

\begin{figure}[tp]
\centering
\includegraphics[width=0.48\columnwidth]{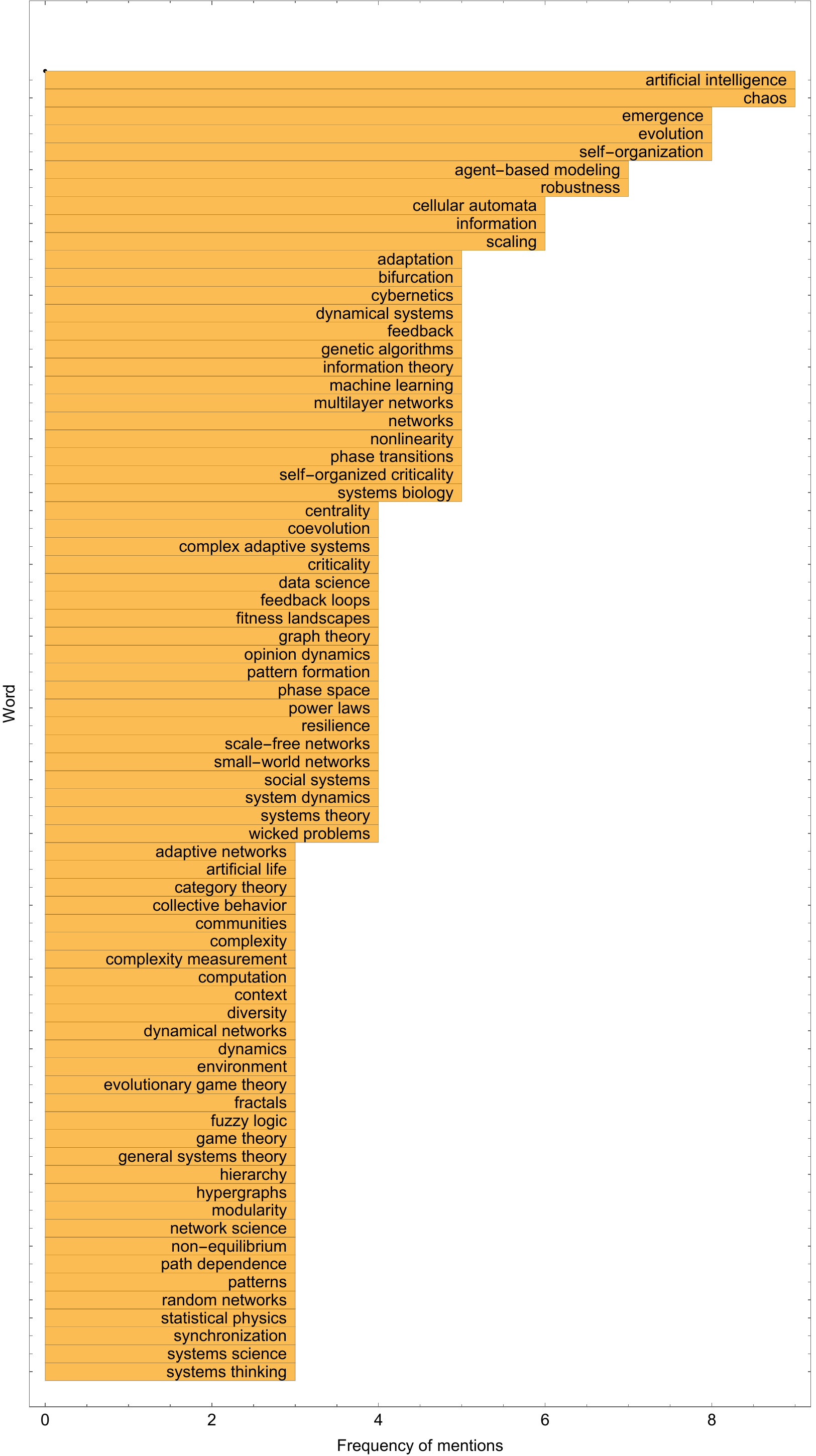} \quad
\includegraphics[width=0.48\columnwidth]{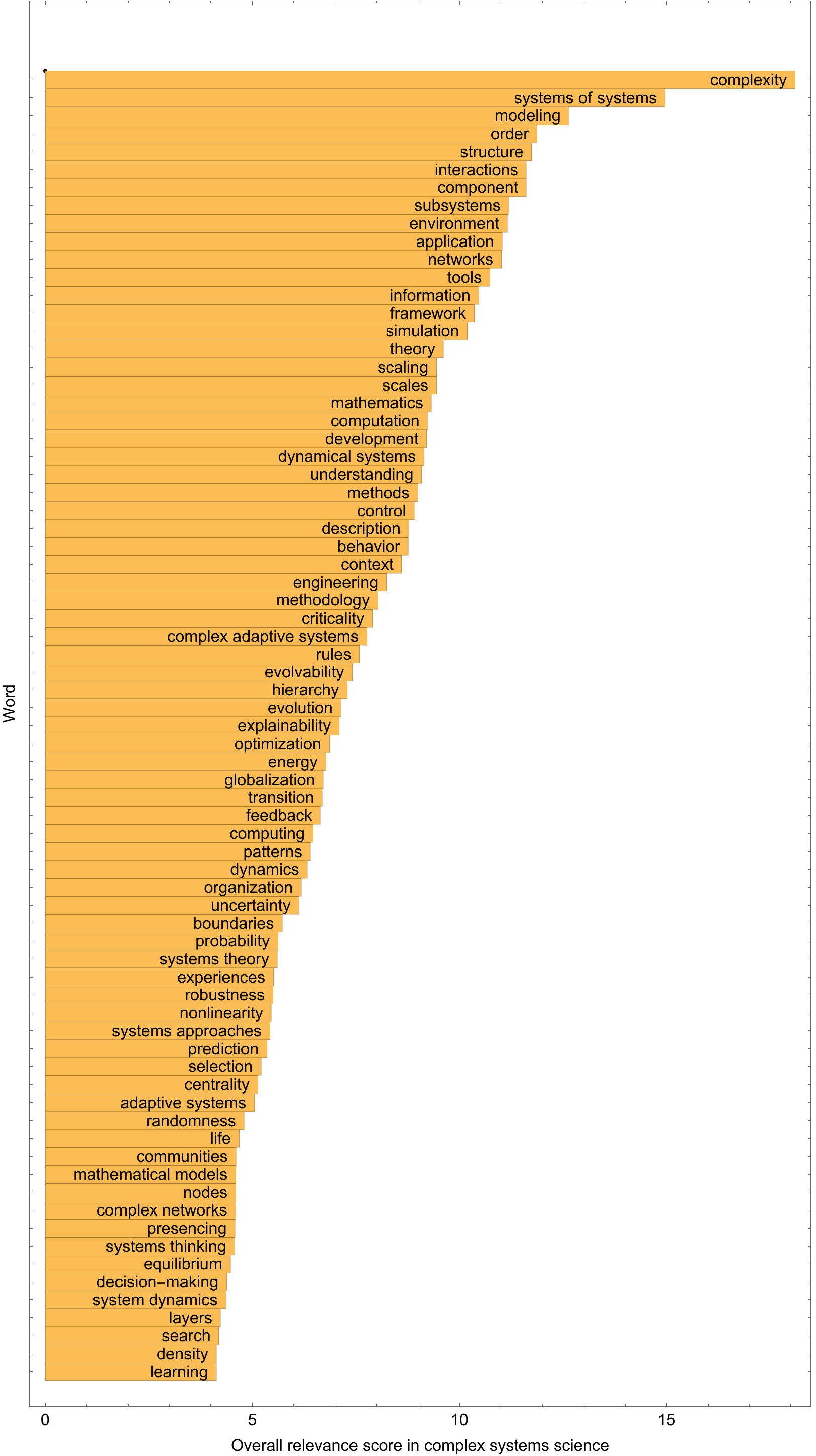}\\~\\
\includegraphics[width=0.48\columnwidth]{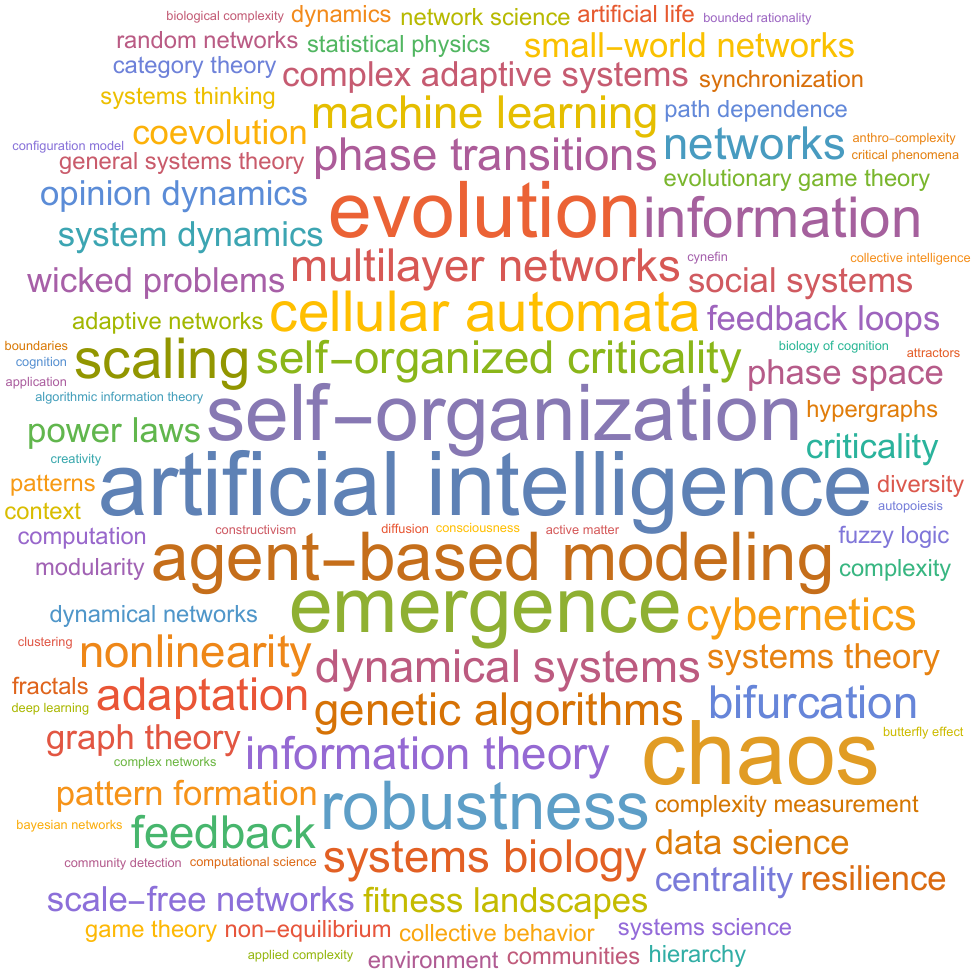} \quad
\includegraphics[width=0.48\columnwidth]{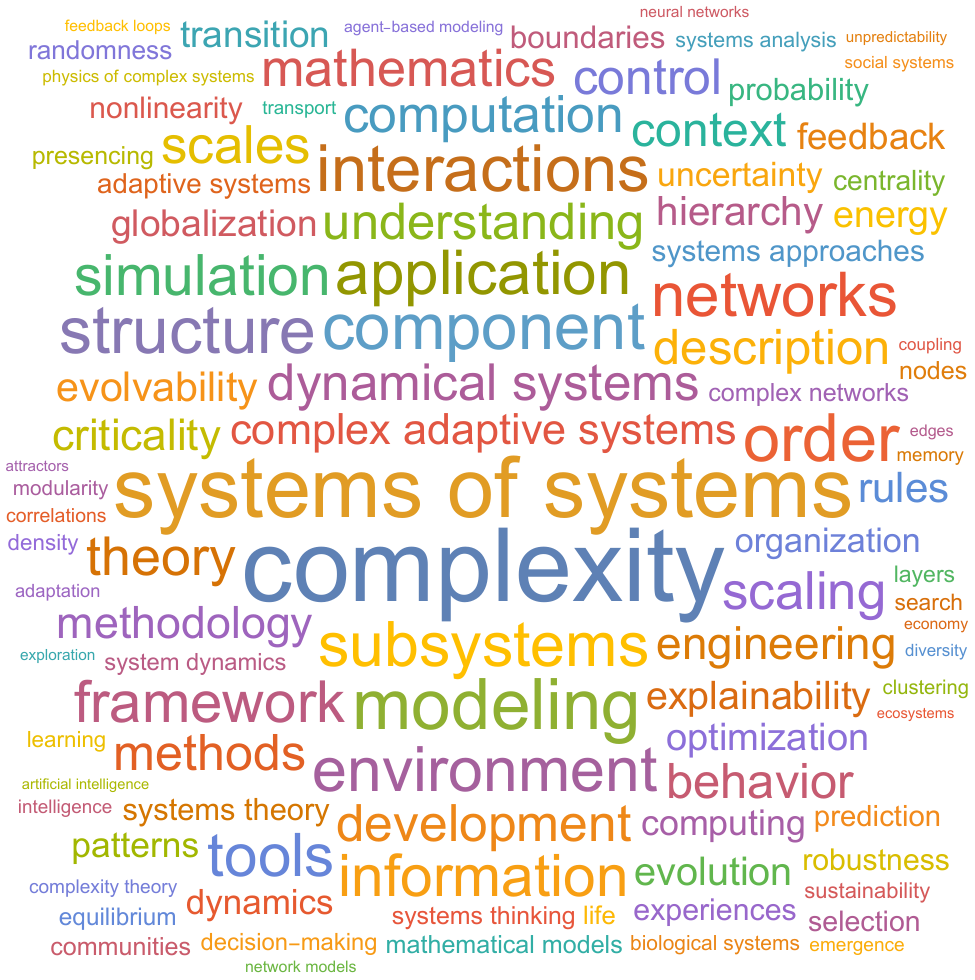}
\caption{Keyword rankings. Left: Keyword ranking based on the frequency of mentions in datasets (a)-(c) (top; only keywords with three or more mentions were listed; a total of 73 keywords) and its word cloud visualization (bottom; including all keywords). Right: Keyword ranking based on the relevance score obtained using dataset (d) (top; only the top 73 keywords) and its word cloud visualization (bottom; including all keywords).}
\label{fig:rankings}
\end{figure}

Figure \ref{fig:rankings} shows the keyword rankings of most frequently mentioned (left) and most relevant (right) keywords, together with their respective word cloud visualizations (bottom). It is visible in these figures that there are interesting differences between the frequencies of mentions (i.e., perception of the community and the public) and the relevance scores obtained using OpenAlex (i.e., actual use in scientific publications). For example, the frequently mentioned keywords such as ``artificial intelligence'', ``chaos'', ``emergence'', ``self-organization'', ``agent-based modeling'', ``cellular automata'', ``bifurcation'' and ``cybernetics'', which are often associated with complex systems in public-facing scientific communications, did not appear in the ranking based on the overall relevance scores. This indicates that the narratives historically used to describe complex systems may not accurately reflect topics and concepts actually discussed with regard to complex systems in scientific literature, justifying the need for the keyword diagram update. Meanwhile, other frequently mentioned keywords such as ``evolution'', ``robustness'', ``information'', ``scaling'', ``dynamical systems'' and ``networks'', were also ranked highly in the relevance score-based ranking, suggesting that those keywords are among the actual key themes of complex systems.

\begin{figure}[t!]
\centering
\includegraphics[width=\columnwidth]{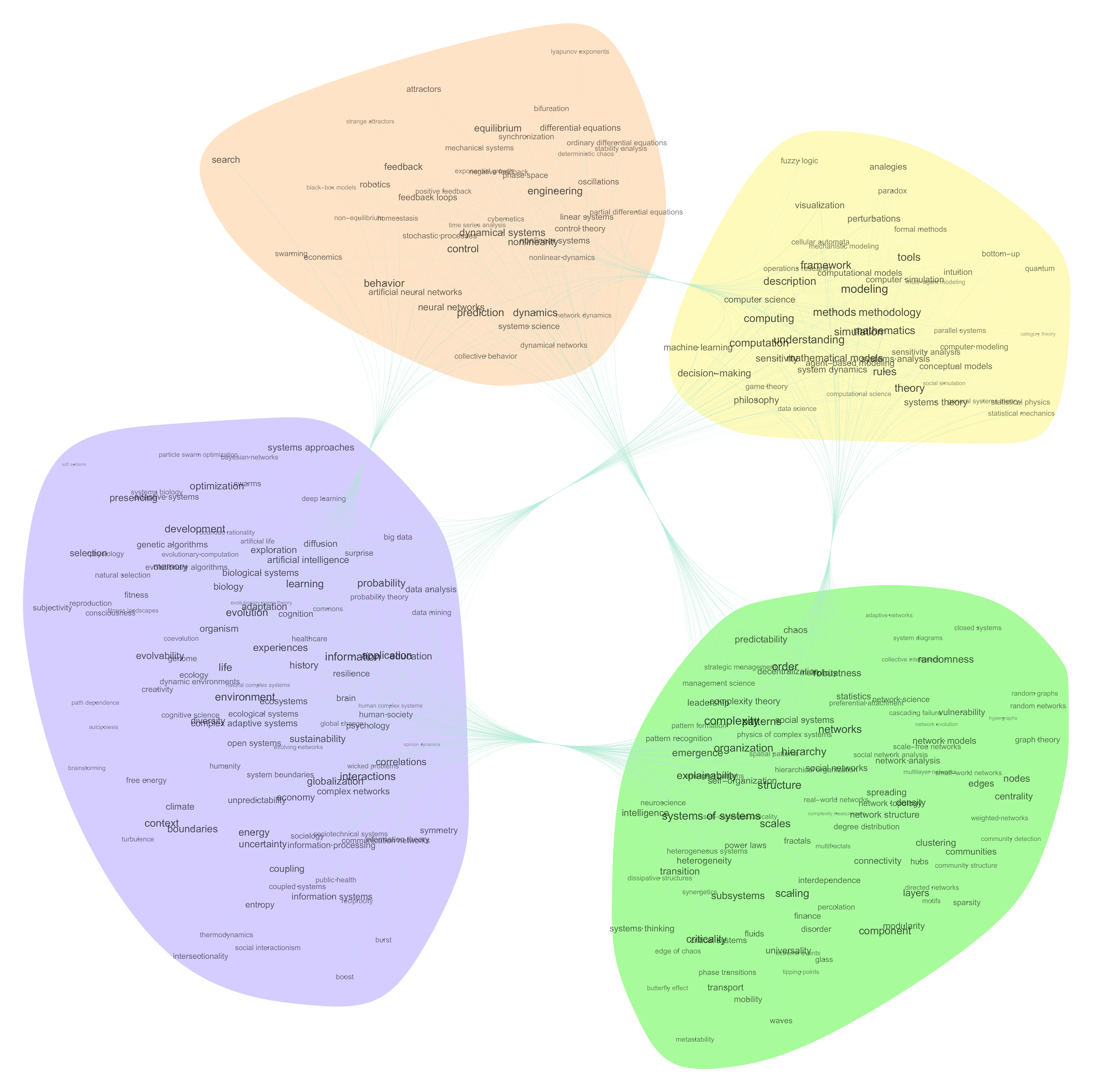}
\caption{Community structure detected in the keyword association network using the Louvain modularity maximization method. Four communities were detected, which can roughly be interpreted as: [Orange, top left] nonlinear dynamics, [Yellow, top right] computational modeling, [Purple, bottom left] biological/ecological/evolutionary/learning/social systems, and [Green, bottom right] networks and systems.}
\label{fig:communities}
\end{figure}

To identify possible topical clusters, the Louvain modularity maximization method \cite{louvain} was applied to the keyword association network, which resulted in four communities (Fig.~\ref{fig:communities}). Detailed inspection of the topics (nodes) categorized in each community resulted in the following labeling as possible interpretations:
\begin{description}
\item[Orange] Nonlinear dynamics
\item[Yellow] Computational modeling
\item[Purple] Biological/ecological/evolutionary/learning/social systems
\item[Green] Networks and systems
\end{description}
These communities may partly reflect the existing traditional disciplinary structure, such as physics/engineering (orange), computer/computational sciences (yellow), mathematical sciences (green), and other more application-oriented fields (purple).

\begin{figure}[t!]
\centering
\includegraphics[width=\columnwidth]{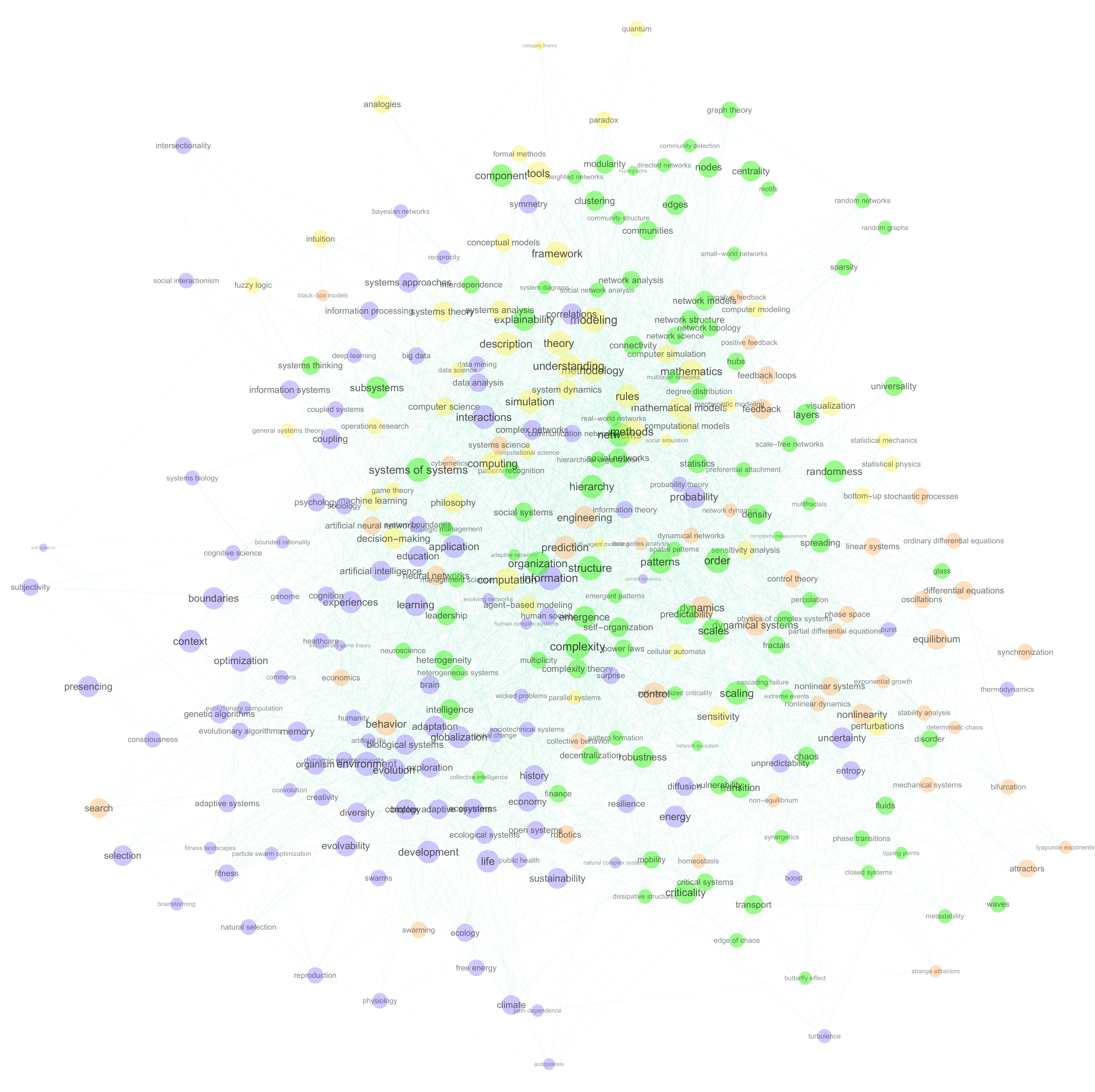}
\caption{Keyword association network visualized using Mathematica's spring electrical embedding layout algorithm. Nodes are colored using the same colors as in Fig.~\ref{fig:communities} according to their community memberships. It is clear that the four ``communities'' detected are actually highly intertwined with each other with no clear separation.}
\label{fig:network}
\end{figure}

However, note that these community labels are no more than just to aid the visibility and organization of the diagram, and they are in fact heavily intertwined with each other if the entire keyword association network is visualized using a conventional network layout algorithm, which does not show any immediately visible community structure (Fig.~\ref{fig:network}). In this entire network visualization, the keywords that form the conceptual core of complex systems naturally come at the center, such as ``complexity'', ``emergence'', ``self-organization'', ``information'', ``structure'', and ``patterns'', even though some of them may not have gained a high relevance score in the literature. The broad range of different topical areas spread from the central core, covering biology, engineering, mathematics, physics, philosophy and others. Readers are encouraged to zoom into this diagram to explore the details and how the topics are placed relative to others in this new keyword diagram.

\begin{figure}[t!]
\centering
\includegraphics[width=\columnwidth]{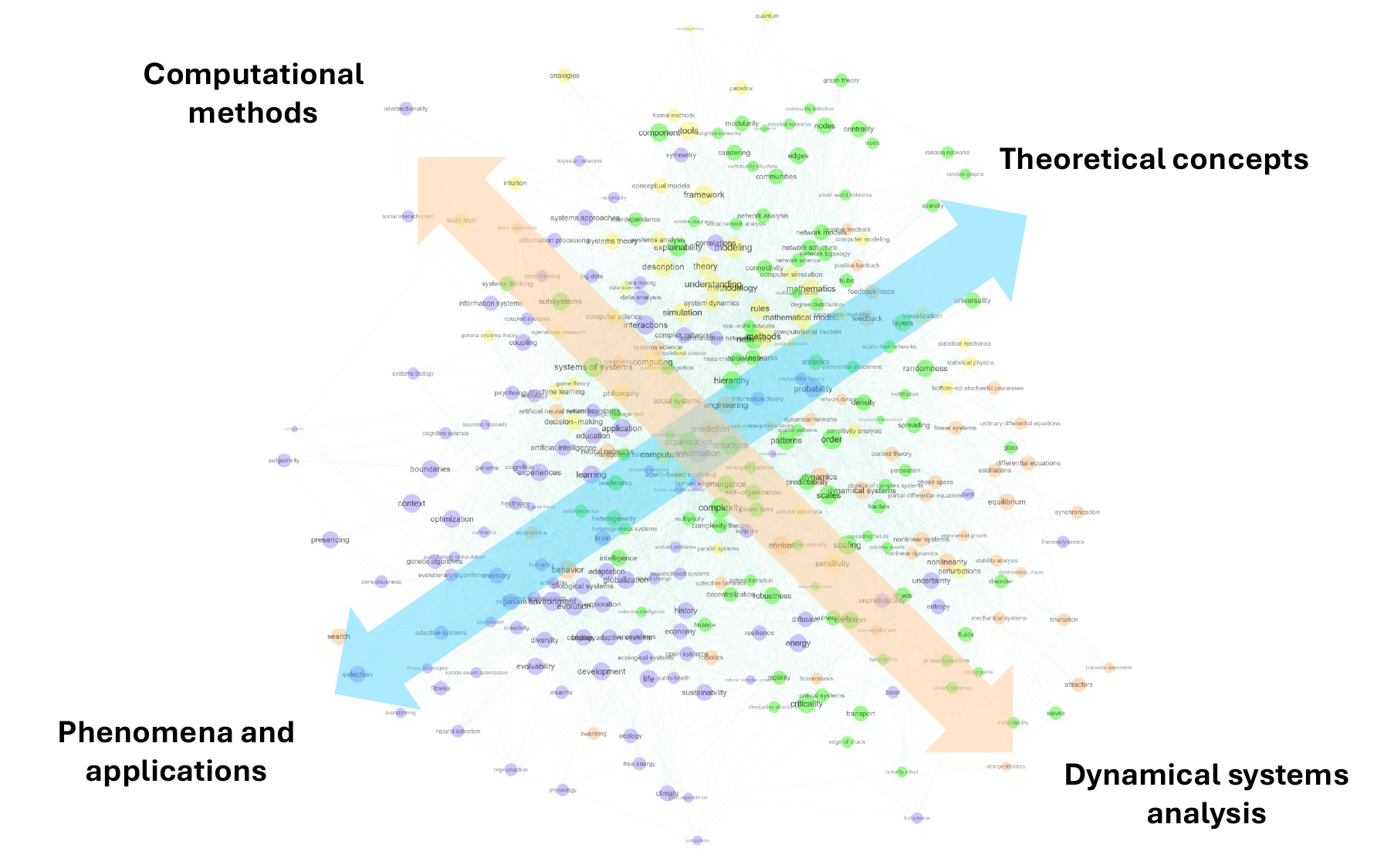}
\caption{An interpretation of large-scale gradients of complex systems topics in the keyword association network. One dimension is the gradient between theoretical concepts and actual phenomena/applications (cyan arrow). The other is the gradient between computational methods and dynamical systems analysis (orange arrow). These dimensions are just suggested interpretation of the overall trends and some topics (nodes) may not follow this interpretation.}
\label{fig:trends}
\end{figure}

An informal yet intuitive observation one can gain from Fig.~\ref{fig:network} is that, even though the four communities detected in Fig.~\ref{fig:communities} are not clearly separated in the actual network, they tend to occupy certain local areas in the network visualization space. Specifically, the purple ``Biological/ecological/evolutionary/learning/social systems'' (applications) cluster tends to spread toward the southwest area of the diagram, while the green ``Networks and systems'' (mathematical sciences) cluster tends to spread toward the opposite direction. Similarly, the yellow ``Computational modeling'' (computer/computational sciences) cluster tends to sit mostly in the northern part of the diagram, whereas the orange ``Nonlinear dynamics'' (physics/engineering) cluster tends to spread more toward the southeast area of the diagram. These observations of the macro-level community distributions, together with more in-depth inspections of locally nearby topics, offers a possible interpretation of the large-scale gradients of complex systems topics in this new diagram, as depicted in Fig.~\ref{fig:trends}. Namely, there appears to be a major dimension of the topical gradient between theoretical concepts and actual phenomena/applications (Fig.~\ref{fig:trends}, cyan arrow), and another between computational methods and dynamical systems analysis (Fig.~\ref{fig:trends}, orange arrow). These dimensions are no more than a crude interpretation of complex high-dimensional interrelationships among the topics, yet they may provide a useful perspective in exploring diverse research topics studied and discussed in the field of complex systems.

\section{Conclusions}

In this study, the keyword diagram of complex systems has been updated in the form of a complex weighted network of associated keywords by using the collective feedback from the community/public and the data obtained from the OpenAlex scientific literature database. The comparison between the data obtained from these two data sources highlighted several topical differences, indicating that there might be some gaps between public perception and actual usage of keywords in scientific publications on complex systems. Such comparison of word uses between public discourse and scientific publications can be a general research methodology potentially useful for other similar topic mapping studies too. Meanwhile, the visualization of the entire association network of keywords did place frequently mentioned keywords near the center of the network (despite their relatively lower relevance scores), implying that those central keywords do play an important role in connecting various other keywords together even in the scientific publications.

The attempt to find topical clusters using the Louvain method produced four communities, but they were all heavily intermingled with each other, resulting in a more nuanced, networked view of the complex systems keywords. Nonetheless, these results of community detection helped recognize the large-scale topical dimensions in the generated keyword diagram that consist of the theory-application axis and the computational-analytical axis. It is our hope that this networked keyword diagram represents the actual state of the field of complex systems science and the breadth and depth of its topics more accurately than the original diagram that might have been overly simplistic and hierarchical.

There are still many limitations and possible future directions in this work. One critical problem is that it still used manually curated keyword associations that would have been heavily influenced by the author's subjective evaluation and filtering. It remains an open question how to select scientifically meaningful associations between keywords solely from the literature data without manual curation. Another major problem is that the two topical axes proposed in this study were simply based on crude interpretation of a network visualization that was made using a specific heuristic network layout algorithm. It would be interesting and more convincing to see if similar axes could be extracted more analytically using a more rigorous dimensionality reduction method. Furthermore, yet another problem is the lack of comparison of the obtained result with results of other similar recent efforts of complex systems topic mapping (e.g., \cite{araasjo,nicholas-project}). It would be interesting and useful to integrate the results of all of those related projects to build a meta-level keyword map, hopefully as a community effort coordinated on a regular basis so that we can track how the relevant topics in complex systems change over time. Finally, exploration of a complex topic network would be extremely difficult on a static visualization and hence it may not carry the same level of pedagogical values as the original 2010 diagram had. It would be much more valuable and useful to have an interactive, dynamic interface by which users can actively explore the relevant keywords and organize the topics into categories in their preferred ways, which is currently under development.

\section*{Acknowledgments}

The author thanks all the community members who provided valuable inputs and all the authors of the resources from which data were collected. The author also thanks Alexandre Guillet for his helpful comments on the initial version of this manuscript. This work was supported in part by the Visitors Program of the Max Planck Institute for the Physics of Complex Systems, Dresden, Germany.


\begin{thebibliography}{99}

\bibitem{ref1} Sayama, H. (2010). Complex systems organizational map. \url{https://en.wikipedia.org/wiki/File:Complex_systems_organizational_map.jpg}.

\bibitem{ref2} Sayama, H. (2022). Posts on several social media, e.g., \url{https://x.com/HirokiSayama/status/1605664684962619394}.

\bibitem{ref3} Mitchell, M. (2009). {\em Complexity: A Guided Tour.} Oxford University Press.

\bibitem{ref4} Barab\'{a}si, A.-L. (2015). {\em Network Science.} Cambridge University Press.

\bibitem{ref5} Sayama, H. (2015). {\em Introduction to the Modeling and Analysis of Complex Systems.} Open SUNY Textbooks.

\bibitem{ref6} Thurner, S., Hanel, R. \& Klimek, P. (2018). {\em Introduction to the Theory of Complex Systems.} Oxford University Press.

\bibitem{ref7} Bar-Yam, Y. (2019). {\em Dynamics of Complex Systems.} CRC Press.

\bibitem{ref8} Menczer, F., Fortunato, S. \& Davis, C. A. (2020). {\em A First Course in Network Science.} Cambridge University Press.

\bibitem{ref9} Ladyman, J. \& Wiesner, K. (2020). {\em What Is a Complex System?} Yale University Press.

\bibitem{ref10} Wikipedia article on ``Complex system''. \url{https://en.wikipedia.org/wiki/Complex_system}.

\bibitem{ref11} De Domenico et al. (2019). {\em Complexity Explained.} \url{https://complexityexplained.github.io/}.

\bibitem{ref12} Castellani, B. \& Gerrits, L. (2021). The 2021 map of the complexity sciences. \url{https://www.art-sciencefactory.com/complexity-map_feb09.html}.

\bibitem{ref13} CCS 2024. (2024). Conference main topics and emerging themes. \url{https://ccs24.cssociety.org/main-topics/}.

\bibitem{ref14} Priem, J., Piwowar, H. \& Orr, R. (2022). OpenAlex: A fully-open index of scholarly works, authors, venues, institutions, and concepts. arXiv preprint arXiv:2205.01833. \url{https://openalex.org/}.

\bibitem{ref15} Sayama, H. \& Akaishi, J. (2012). Characterizing interdisciplinarity of researchers and research topics using web search engines. {\em PLOS ONE}, 7(6), e38747.

\bibitem{project-github} Sayama, H. (2024). Complex systems topic network GitHub page. \url{https://github.com/hsayama/complex-systems-topic-network-2024}.

\bibitem{louvain} Blondel, V. D., Guillaume, J. L., Lambiotte, R., \& Lefebvre, E. (2008). Fast unfolding of communities in large networks. {\em Journal of Statistical Mechanics: Theory and Experiment}, 2008(10), P10008.

\bibitem{araasjo} Araãšjo, T., Abreu, A., \& Louã, F. (2024). The evolution of complexity co-occurring keywords: Bibliometric analysis and network approach. {\em Advances in Complex Systems}, 27(04n05), 1-23.

\bibitem{nicholas-project} Nicolás-Carlock, J. R. (2025). ``Complexity: 5 Questions'' bipartite network project. \url{https://github.com/jrncarlock/research-supporting-data/tree/main/complexity}.


\end{thebibliography}
\end{document}